\documentclass[%
 reprint, 
 superscriptaddress,
 amsmath,amssymb,
 aps,
 prl,
]{revtex4-2}

\usepackage[version=3]{mhchem} 
\usepackage[export]{adjustbox}
\usepackage{comment}
\usepackage{etoolbox}
\usepackage{amsfonts} 
\usepackage{amssymb} 
\usepackage{svg}
\usepackage[utf8]{inputenc}
\usepackage[T1]{fontenc}
\usepackage{textcomp}

\newif\ifcomment 
\commenttrue 

\ifcomment
  \includecomment{comment}
\else
  \excludecomment{comment} 
\fi

\newcommand*\lp[1]{}
\newcommand*\jlh[1]{}
\newcommand*\superfluous[1]{}
\newcommand*\hemicylinders[1]{#1}
\newcommand*\summary[1]{}

\ifcsdef{achemso}{
    \author{Johannes L. H\"ormann}
    \author{Lars Pastewka}
    \affiliation[University of Freiburg]
    {IMTEK, University of Freiburg, Freiburg i. Br., Germany}
}
\ifcsdef{achemso}{
    \title[An \textsf{achemso} demo]
      {SDS on Au(111)}
}{}
\ifcsdef{abbreviations}{
    \abbreviations{IR,NMR,UV}
    \fi
}
\newcommand{\abstracttext}{
We use molecular dynamics simulations to study the frictional response of the anionic surfactant sodium dodecyl sulfate (SDS) monolayers and hemicylindrical aggregates physisorbed on gold.
Our simulations of a sliding spherical asperity reveals two friction regimes: At low loads, the films show Amontons' friction with a friction force that rises linearly with normal load.
At high loads, the friction force is independent of load as long as no direct solid-solid contact occurs.
The transition between these two regimes happens when only a single molecular layer is confined in the gap between the sliding bodies.
The friction force at high loads on a monolayer rises monotonically with film density and drops slightly with the transition to hemicylindrical aggregates.
This monotonous increase of friction force is compatible with a traditional plowing model of sliding friction.
At low loads, the friction coefficient reaches a minimum at intermediate surface concentrations.
We attribute this behavior to a competition between adhesive forces, repulsion of the compressed film, and the onset of plowing.
}

\begin{document}

\ifcsdef{achemso}{
    \begin{tocentry}
    
    Some journals require a graphical entry for the Table of Contents.
    This should be laid out ``print ready'' so that the sizing of the
    text is correct.
    
    Inside the \texttt{tocentry} environment, the font used is Helvetica
    8\,pt, as required by \emph{Journal of the American Chemical
    Society}.
    
    The surrounding frame is 9\,cm by 3.5\,cm, which is the maximum
    permitted for  \emph{Journal of the American Chemical Society}
    graphical table of content entries. The box will not resize if the
    content is too big: instead it will overflow the edge of the box.
    
    This box and the associated title will always be printed on a
    separate page at the end of the document.
    
    \end{tocentry}
}
\ifcsdef{achemso}{}{
    \title{Molecular simulations of sliding on SDS surfactant films}
    \author{Johannes L. H\"ormann}
    \email{johannes.hoermann@imtek.uni-freiburg.de}
    \affiliation{Department of Microsystems Engineering, University of Freiburg, Georges-K\"ohler-Allee 103, 79110 Freiburg, Germany}
    \affiliation{Cluster of Excellence livMatS, Freiburg Center for Interactive Materials and Bioinspired Technologies, University of Freiburg, Georges-K\"ohler-Allee 105, 79110 Freiburg, Germany}
    \author{Chenxu Liu}
    \author{Yonggang Meng}
    \affiliation{State Key Laboratory of Tribology in Advanced Equipment, Lee Shau Kee Science and Technology Building, Tsinghua University, Haidian District, Beijing, 100084, China}
    \author{Lars Pastewka}
    \affiliation{Department of Microsystems Engineering, University of Freiburg, Georges-K\"ohler-Allee 103, 79110 Freiburg, Germany}
    \affiliation{Cluster of Excellence livMatS, Freiburg Center for Interactive Materials and Bioinspired Technologies, University of Freiburg, Georges-K\"ohler-Allee 105, 79110 Freiburg, Germany}
    \begin{abstract}
        \abstracttext    
    \end{abstract}
    \maketitle
}

\ifcsdef{achemso}{
    \begin{abstract}
      \abstracttext
    \end{abstract}
}

\lp{Comments by Lars Pastewka}

\jlh{Comments by Johannes Hörmann}

\superfluous{Suggested for removal}

\section{Introduction}
\label{sec:introduction}

\begin{figure}
  \includegraphics[width=240pt]{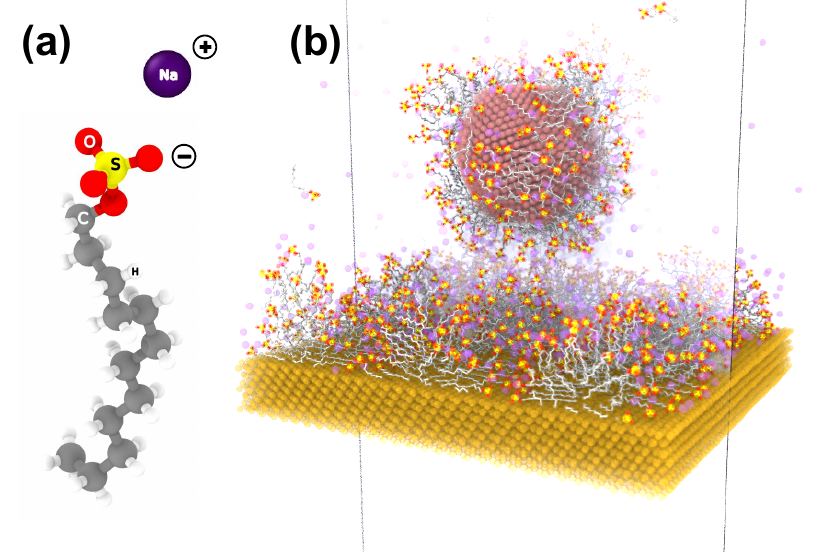}
  \caption{\textbf{(a)} The DS\textsuperscript{-} anion with hydrophilic sulfate head group and hydrophobic hydrocarbon tail and the Na\textsuperscript{+} counterion making up SDS. \textbf{(b)} Perspective view on probe model above hemicylindrical SDS aggregates on substrate with cross-section plane indicated.}
  \label{fig:perspective-view}
\end{figure}

In aqueous solution, amphiphilic surfactants adsorb on immersed surfaces.
Depending on the specific molecule and concentration, they form surface structures of varying morphology: monolayers of flat-lying molecules, hemicylindrical stripes or full cylinders have all been observed.
The anionic model surfactant sodium dodecyl sulfate (SDS, see Fig.~\ref{fig:perspective-view}a) provides both molecular simplicity and a rich adsorption film phase diagram with transition from monolayers to hemicylinders~\cite{jaschke1997surfactant,chen2009potential}.
Mechanical properties of SDS films depend on adsorbate concentration~\cite{burgess1999direct,zhang2015boundary}.
Even macroscopic frictional properties appear to be influenced by concentration, and it has been hypothesized that this is related to the morphological transition in these films~\cite{zhang2015boundary}.
Since for ionic surfactants, adsorption and film phase morphology are controllable electrochemically, SDS adsorption films have become a model system for the study of electrotunable friction and lubrication~\cite{yang2014potentialcontrolled,zhang2014stick,zhang2015boundary,zhang2015effect,zhang2015control,liu2021active,liu2022online,liu2023mitigation}.

\begin{figure*}
  \includegraphics[width=400pt]{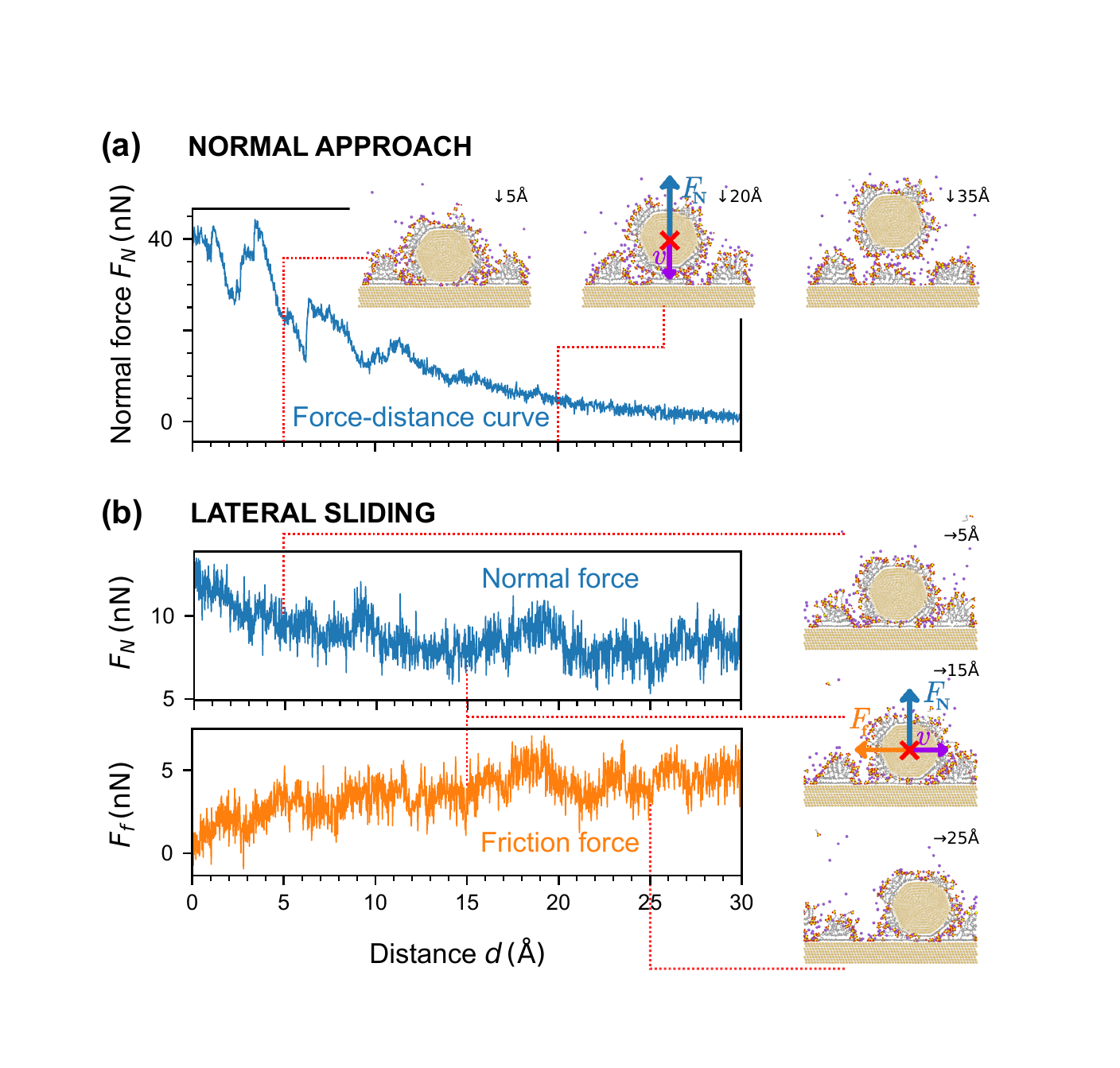}
  \caption{\textbf{(a)} Normal approach force-distance curve at
  $1\,\mathrm{m}\,\mathrm{s}^{-1}$ with snapshots at $35$, $20$, and
  $5\,\mathrm{\text{\AA}}$ probe-substrate surface-surface distance.
  \textbf{(b)} Lateral sliding at $1\,\mathrm{m}\,\mathrm{s}^{-1}$
  normal and friction forces with snapshots at $5$, $15$, and
  $25\,\mathrm{\text{\AA}}$ lateral sliding distance and fixed normal
  probe-substrate surface-surface distance of $9\,\mathrm{\text{\AA}}$.}
  \label{fig:concept-afm-approach}
\end{figure*}

In this work, we use molecular calculations to probe SDS adsorption films at the Au(111)-water interface with a nanoscale sphere (see Fig.~\ref{fig:perspective-view}b), that can be representative of either an atomic force microscope (AFM) tip or an asperity on a rough surface.
Specifically, we record force-distance curves on normal approach of a model AFM probe and friction force measurements on lateral sliding of the same probe on SDS adsorption films at the aqueous solution-Au(111) interface as conceptually shown in Fig.~\ref{fig:concept-afm-approach}. 
We systematically explore the parametric space spanned by concentration, film morphology and normal force.
We find that single-asperity friction on these films has two distinct regions: At low loads, the friction force rises with normal force before saturating.
Both, the friction coefficient at low loads and the saturation force depend on film concentration.
%

SDS has been a model surfactant of colloidal science at least since the mid 20th century -- a simple molecule with a complex phase diagram.
Early experimental studies focused on the properties of SDS micelles, which self-assemble at the critical micelle concentration (CMC) of about $8.2\,\mathrm{mM}$~\cite{mukerjee1971critical}.
At high concentration of about $60~\text{mM}$~\cite{miura1972seconda, kodama1972secondb, kodama1972secondc, kubota1973second}, there is a transition to rod-like structures~\cite{bezzobotnov1988temperature,kekicheff1989phase,bales1998precision}.
The number of molecules in the micelle, the aggregation number, appears to be around $\sim 60$ molecules and independent of concentration above the CMC but below the transition to rods~\cite{turro1978luminescent}.
The whole phase diagram is more complicated, as micelles and rod can themselves self-assemble into crystalline or partially crystalline structures.


Micellization in bulk solution and the adsorption film phase transition to hemicylindrical aggregates at the interface arise due to the same hydrophobicity-induced self-organization mechanism.
SDS adsorption has been investigated for numerous idealized solid-solution systems, such as in aqueous solution on graphite~\cite{wanless1996organization,wanless1997surface,wanless1997weak}, aluminum~\cite{karlsson2008adsorption}, or gold~\cite{jaschke1997surfactant,burgess1999direct,burgess2001electrochemical,petri2002nanostructuring,soares2007sodium,chen2009potential}.
Jaschke et al.~\cite{jaschke1997surfactant} first showed direct atomic-force microscopy (AFM) images of stripe-like SDS adsorption aggregates on flat Au(111) and interpreted these as hemicylinders.
They measured a spacing of $4.9\pm 0.5\,\mathrm{nm}$ between the stripes' central axes, and the same structures have also been observed on rough surfaces~\cite{schniepp2007surfactant}.
In the realm of electrochemistry, Burgess et al.~\cite{burgess1999direct} showed a potential-controlled transition from hemicylindrical to condensed adsorption film phase for SDS on flat Au(111) (with an aggregate periodicity of $4.4\pm0.5\,\mathrm{nm}$).
Their investigation was later extended onto a broad range of electrode potentials and SDS concentrations~\cite{burgess2001electrochemical}.
The parametric space spanned by electrode potential and surfactant concentration gives rise to at least four distinct adsorption film phases, namely complete desorption, flat monolayers, hemicylindrical aggregates, and possibly a compact bilayer phase~\cite{chen2009potential}.
SDS aggregation on other substrates, e.g. on stainless steel~\cite{he2011correlation}, graphite~\cite{wanless1996organization, wanless1997surface}, or oxidized graphene~\cite{glover2012chargedriven}, shows identical morphological transitions with concentration.


Film phase transitions may mean changing mechanical properties.
Tribologists have investigated the boundary lubrication performance of SDS adsorption films, in particular with a focus on the possibilities for electrotunable friction in macroscopic tribotests on stainless steel.
%
%
Zhang and Meng~\cite{zhang2014stick, zhang2015boundary} carried out sphere-on-disk experiments with a macroscopic zirconium dioxide probe of radius $\sim 6\,\text{mm}$.
They observed a change in the behavior across the transition from flat-lying monomers to hemicylindrical aggregates with increasing surfactant concentration.
In their series of experiments, the mean friction coefficient had a minimum at an intermediate SDS concentration, where surface aggregates consisted of densely packed monolayers.
%
%
%
This appears to indicate a morphology-dependent change in the boundary lubrication behavior.

Zhang and Meng~\cite{zhang2014stick} argued that compared to dense films, stripe-like aggregates expose the metallic bulk between the aggregates, leading to an increased likelihood of metal-metal junction formation.
Yet, whether this observation is actually a consequence of morphology remains unclear.
In particular, the direct contribution of film morphology at molecular scales as e.g. probed in AFM experiments is not understood. 
Hence, we here model AFM experiments on SDS adsorption films at the \ce{H2O - Au (111)} interface by means of classical all-atom (AA) molecular dynamics (MD).


Important related systems studied by the friction community are organic friction modifiers (OFMs), a class of non-ionic surfactants, in oil. OFM adsorption films are referred to as self-assembled monolayers (SAMs)~\cite{knippenberg2008atomic,chandross2008simulations}.
Nonequilibrium simulations of OFMs~\cite{ewen2016nonequilibrium,ewen2018advances} of such films are often studied in parallel-plate geometries, that have a constant film thickness and cannot describe plowing.
More complex geometries, such as crossed cylinders~\cite{debeer2013alternative,debeer2014solventinduced} or tips~\cite{gao2021scaledependent} allow insights into the reorientation of the adsorbant molecules but require larger simulations.
We here employ a sphere-on-flat geometry, since we believe that plowing the SDS film is a crucial friction mechanism.

\section{Methods}
\label{sec:methods}

%
The prediction of adsorption isotherms for complex ionic molecules is presently out-of-reach for molecular calculations.
To connect our molecular systems to experiments, we therefore measured the adsorption behavior of SDS molecules on the interface of an Au coated quartz crystal sensor and SDS aqueous solution ($0.01\sim20\,\mathrm{mM}$) with a quartz crystal microbalance (QCM, Q-sense E4 system, Biolin Scientific, Sweden). 
The changes of both resonance frequency ($\Delta f$) and dissipation factor ($\Delta D$) due to the mass variations of ions or molecules on the sensors per unit area were recorded by the QCM.
Initially, we started the measurement in air to find the base resonance frequencies until approaching stable frequency values.
Then, we injected pure water into the testing module by an external pump with the speed of $200\,\mathrm{\mu L}\,\mathrm{min}^{-1}$ through polytetrafluoroethylene (PTFE) tubing, monitoring the fluctuations of $\Delta D$ and $\Delta f$ in water until they stabilize.
Finally, we switched the injection from pure water to the SDS solution, and recorded the changes in $\Delta D$ and $\Delta f$ caused by adsorption or desorption of the ions or molecules for the third, fifth, seventh, and ninth overtones at $298\,\mathrm{K}$.
When the measured value of $\Delta D$ falls below $10^{-6}$, we regard the adsorbed SDS as rigid and use the Sauerbrey equation 
$\Delta m = -C_0 \Delta f/N$
to calculate the adsorbed mass changes $\Delta m$ from the frequency changes at the $N$th overtone of the oscillations~\cite{sauerbrey1959verwendung,thavorn2014competitive,zhang2015effect}.
The \emph{mass sensitivity constant} $C_0$ arises from intrinsic material properties of the quartz crystal~\cite{sauerbrey1959verwendung}. In this work, we use $C_0 = 17.7\,\mathrm{ng}\,\mathrm{Hz}^{-1}\,\mathrm{cm}^{-2}$ for an AT-cut crystal of fundamental resonance frequency $f_0 = 5\,\mathrm{MHz}$~\cite[p. 201]{heising1946quartz} and the third overtone for the calculation because of its better signal-to-noise ratio among the different overtones.

%
At selected configurations along this experimental adsorption isotherm, we carry out classical MD simulations employing a valence force-field that allows modeling 
of physisorption processes but does not include breaking of covalent bonds. 
In the context of friction we care about the explicit description of hydrogen atoms, because coarse-grained models systematically underestimate frictional dissipation in non-equilibrium calculations~\cite{ewen2016comparison}.
Among all-atom parametrizations, the CHARMM branch has been established in a careful series of iterations~\cite{roux1990theoretical,mackerell1995allatom,schlenkrich1996empirical,mackerell1998allatom} and used successfully for modeling micelles in bulk by several independent groups~\cite{yoshii2006molecular,yoshii2006moleculara,yoshii2006molecularb,tang2014molecular}.
CHARMM36 can be combined with a parametrization for the interaction of molecules with solids including Au through the INTERFACE force-field~\cite{heinz2013thermodynamically}.
Hence we use the CHARMM36 force-field~\cite{best2012optimization} and the rigid water model TIP3P (CHARMM standard) as an explicit solvent.
An embedded atom method potential by Grochola~\cite{grochola2005fitting} describes the Au-Au interaction.
All simulations run at standard conditions of temperature $T = 298\,\text{K}$ and pressure $P = 1013\,\text{hPa}$.
Except for initial thermalization, we use a Galilean-invariant dissipative particle dynamics (DPD) thermostat~\cite{groot1997dissipative,soddemann2003dissipative} applied only to probe and substrate (see below) but not solvent and salt. 

All simulated systems consist of a model AFM probe of $5\,\mathrm{nm}$ diameter 
and substrate block of roughly $15\,\mathrm{nm}\times 15\,\mathrm{nm}\times 15\,\mathrm{nm}$ in size. 
Probe and substrate are composed of uncharged gold, with outer (111) planes facing each other. 
AFM probe models are prepared by melting gold spheres of 
3873 atoms and $2.5\: \text{nm}$ initial 
radius and subsequent slow quenching
from $1800\: \text{K}$ down to $298\: \text{K}$ over 
a time span of $100\: \text{ns}$ at $5\: \text{fs}$ time step.
This yields a single crystalline, almost spherical, gold probe shown in Fig.~\ref{fig:perspective-view}b.
The substrate is a cubic, single crystal gold block with one of its $\{111\}$ surfaces exposed 
to the solution.
All systems are solvated in water and carry zero net charge.
SDS adsorption films cover substrate and probe.


The specific size of the substrate was chosen to accommodate three hemicylindrical surfactant 
aggregates of $\sim 2.5\: \text{nm}$ radius~\cite{jaschke1997surfactant}.

Sampling the full adsorption process from solution lies out of reach within
the timescales accessible to brute-force MD~\cite{rai2017modeling}.
While free-energy calculations could in principle yield insights into the absorption process~\cite{yoshii2006moleculara}, we here preassemble a set of monolayers and hemicylinders at appropriate surface concentrations to cover a wide range of the adsorption isotherm across the regime of film phase transition.
Likewise, we assemble a uniform monolayer of equivalent surface concentrations around the
probe model to mimic the double layer structure between probe and substrate.
We arrange sufactant molecules on the surface of the minimal sphere enclosing the gold
cluster and close the remaining gaps between surfactant tail and flat facets by an artificial
pulling force.

Both preassembled probe and substrate systems are solvated and relaxed in separate simulations.
The relaxation procedure starts with an energy minimization, followed by $100\,\mathrm{ps}$ NVT and $100\,\mathrm{ps}$ NPT
equilibration under restrained ion positions.
Those calculations are followed by an unrestrained $3\,\mathrm{ns}$ NPT
relaxation of the preassembled adsorption layer.
%
Probe and substrate are merged at an initial distance of $d = 3\,\text{nm}$ -- sufficiently far to not disturb the adsorption film structure.
In the case of overlap, the conflict is solved by removing solvent molecules.
If removing solvent molecules alone cannot resolve the
conflict (i.e. in the case of overlapping ions and substrate atoms), the ions in question are
moved around randomly until the conflict can be resolved by removing solvent molecules only.
The assembled system is then equilibrated according to the same procedure described above.

For normal approach, we set the probe model’s frozen core of $2.4\,\text{nm}$ diameter in instantaneously motion at a prescribed velocity of $0.1$, $1$, or $10\,\mathrm{m}\,\mathrm{s}^{-1}$ as stated hereafter.
The substrate's bottom-most $1.4\,\text{nm}$ layer is also frozen to fix its position.
%
For lateral sliding runs at fixed probe-substrate distances,
we extract equidistant configurations along the normal approach
trajectories.
These individual snapshots are relaxed again for $20\,\mathrm{ps}$.
We then assign a constant lateral velocity of $1\,\mathrm{m}\,\mathrm{s}^{-1}$ to the probe's core.
%
All surface-surface distances are stated with respect to perfect overlap of substrate's and probe's outermost atomic layer  as zero reference.

Our sliding simulations are carried out at constant height of the probe and at fixed velocity $1\,\mathrm{m}\,\mathrm{s}^{-1}$.
In similar studies, friction forces and normal distances are often sampled under constant normal force applied via a spring attached to the probe model~\cite{gao2021scaledependent}.
Our approach avoids the introduction of an additional spring constant, suppressing stick-slip motion~\cite{thompson1990origin} and associated relaxation phenomena.
In the context of experimental realization of our system, the results presented here may be more representative for a single asperity on a rough surface that is embedded into a relatively stiff surrounding medium, rather than an AFM tip that is typically connected to a soft cantilever~\cite{mate2019tribology}.


\section{Results}

\begin{figure}
  \includegraphics[width=240pt]{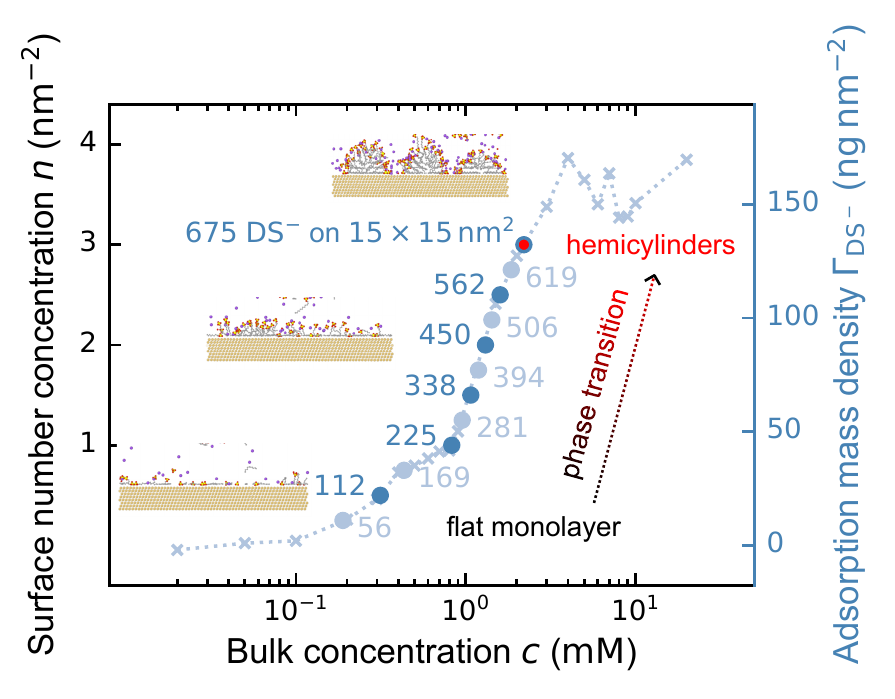}
  \caption{Experimental SDS on Au(111) adsorption isotherm in aqueous solution recorded by quartz crystal microbalance (QCM). Experimental measurements are indicated by "x" markers and connected by dotted line. Labeled circular data points show the actual numbers of $\mathrm{DS}^{-}$ anions assembled on a $15\times15\,\mathrm{nm}^{2}$ gold substrate in this for for sampling representative configurations across the isotherm. Alternating light and strong coloring helps to associate labels and markers.}
  \label{fig:sds_on_au_111_adsorption_isotherm}
\end{figure}

Figure~\ref{fig:sds_on_au_111_adsorption_isotherm} shows the experimentally recorded SDS adsorption isotherm on Au(111) in aqueous solution ranging from $0.25\,\mathrm{nm}^{-2}$ up to $3\,\mathrm{nm}^{-2}$ surface concentration.
Solid circles in Fig.~\ref{fig:sds_on_au_111_adsorption_isotherm} show the concentrations at which MD calculations are carried out.
The labels next to these points give the corresponding number of molecules on a $15 \times 15\,\mathrm{nm}^2$ substrate.
We cover the full concentration range from $0.1\,\mathrm{mM}$ up to saturation at about $3\,\mathrm{mM}$, across which the adsorption film transitions from flat lying monolayers to hemicylindrical aggregates.
At low to intermediate concentrations, samples are preassembled as monolayers.
At high concentration, we preassemble hemicylinders at the red discretization point at the upper concentration end corresponding to $3\,\mathrm{nm}^{-2}$ or $675$ molecules.
These hemicylinders were stable over $50\,\mathrm{ns}$, our longest molecular dynamics run.
We consider this system representative for the plateau region that emerges at high bulk concentration.
All our systems are homogeneous, i.e. either pure monolayers or hemicylinders, while in reality mixed phases of monolayers and hemicylinders exist~\cite{burgess1999direct, chen2009potential}.
%

\begin{figure}
  \includegraphics[width=240pt]{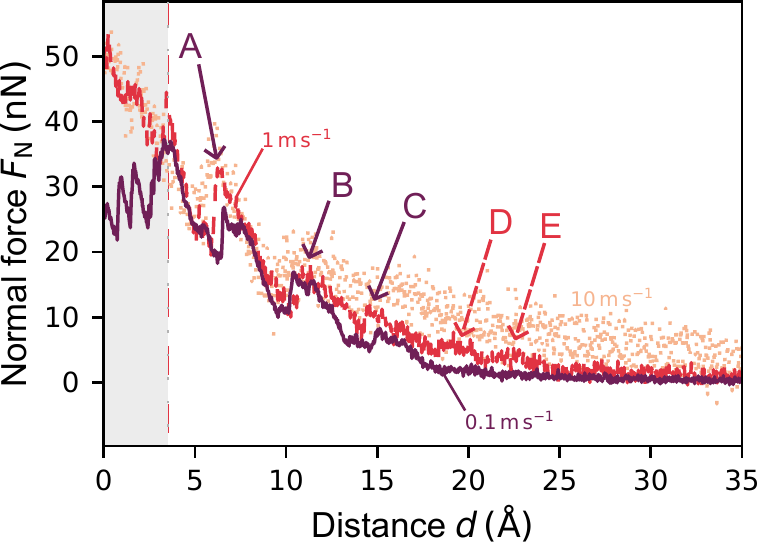}
  \caption{AFM force-distance curves for probe coming down centered above hemicylindrical aggregate at velocities $10\mathrm{m}\,\mathrm{s}^{-1}$ (light, dotted), $1\mathrm{m}\,\mathrm{s}^{-1}$ (dashed), and $0.1\mathrm{m}\,\mathrm{s}^{-1}$ (dark, solid). Shown data points represent $0.2\,\mathrm{ps}$, $2\,\mathrm{ps}$, and $20\,\mathrm{ps}$ averages respectively. Features A-C (dark purple arrows) are qualitatively observable in both $1\mathrm{m}\,\mathrm{s}^{-1}$ and $0.1\mathrm{m}\,\mathrm{s}^{-1}$ trajectories. Features D and E (dashed red arrows) are only observed in the $1\mathrm{m}\,\mathrm{s}^{-1}$ trajectory. The grayed area below $3\,\mathrm{\text{\AA}}$ normal distance  Trajectories begin at $5,\mathrm{nm}$ normal surface-surface distance. We do not show the regime of constant viscous drag beyond $3.5\,\mathrm{nm}$.}
  \label{fig:afm_force_distance_curves_velocity}
\end{figure}

Since we are interested in the effect of adsorption films on force response, we first need to rule out that we are measuring the viscous drag of the solvent in our calculations.
Hence, we estimate viscous drag of our model AFM probe in water at standard conditions
by means of Stokes' law for a bead in viscous fluid, $F_d = 6 \pi \eta r v$,
with probe radius ${r = 2.5\: \text{nm}}$ using the reported
dynamic viscosity of the TIP3P rigid water,
$\eta = 0.321\: \text{mPa}\: \text{s}$~\cite{gonzalez2010shear}.
Additionally, we run probe normal approaches on the system illustrated in 
Fig.~\ref{fig:perspective-view} for the velocities $10\,\mathrm{m}\,\mathrm{s}^{-1}$,
$1\,\mathrm{m}\,\mathrm{s}^{-1}$, and $0.1\,\mathrm{m}\,\mathrm{s}^{-1}$ and show normal force versus distance in Fig.~\ref{fig:afm_force_distance_curves_velocity}.

In our MD results, we interpret average forces acting on the probe at distances $5 \geq d > 4\: \text{nm}$ above the substrate as viscous drag.
Table~\ref{tab:results_stokes_drag} summarizes the analytical and MD estimates.
%
Both  analytical estimate and MD results extracted from force-distance curves yield non-negligible drag for $v \gtrsim 10\: \text{m}\: \text{s}^{-1}$.
Viscous drag at speeds slower than $10\: \text{m}\: \text{s}^{-1}$ is an order of magnitude lower than forces from resistance of the surfactant films.
Deviations from analytical Stokes estimates up to an order of magnitude are
likely due to hydrodynamic interactions between tip and surface at the distances
of $\lesssim 5\,\text{nm}$ probed here, that correspond just to a couple of hydration shells.
The MD drag estimates for the film-covered probe approaching a hemicylindrical aggregate are larger than for the bare probe approaching a monolayer~\cite{hormann2020sds}.

%
In the force-distance curves of Fig.~\ref{fig:afm_force_distance_curves_velocity}, there are at least three quantitatively relatable features marked by dark purple arrows A, B and C  at about $6$, $11$ and $15\,\mathrm{\text{\AA}}$.
The latter two of these features are unidentifiable in the fast approach case of  $10\,\mathrm{m}\,\mathrm{s}^{-1}$. 
Interestingly, the $1\,\mathrm{m}\,\mathrm{s}^{-1}$ approach appears to exhibit two more features marked by red dashed arrows D and E at $19$ and $22\,\mathrm{\text{\AA}}$ that
are not identifiable in the $0.1\,\mathrm{m}\,\mathrm{s}^{-1}$ approach. 
Common to all MD results is the probe-substrate contact at about $4\,\mathrm{\text{\AA}}$ distance and the onset of plastic deformation below, indicated by a gray area. 
Other than the two faster approaches, the $0.1\,\mathrm{m}\,\mathrm{s}^{-1}$
approach velocity exhibits three subsequent deformation events and decreased
repulsion below $3\,\mathrm{\text{\AA}}$ normal distance.

\begin{table}
\footnotesize
\centerline{
{\renewcommand{\arraystretch}{1.2}%
\begin{tabular}{|r|r|r|r|}
\hline
   \multicolumn{4}{|c|} { Stokes drag of the AFM probe } \\
   \hline
   velocity $v\: (\text{m}\,\text{s}^{-1})$  & 10 & 1  & 0.1 \\
   \hline 
   analytical $F_d\,(\text{nN})$     &  0.15 & 0.02 & $\sim$ 0  \\
   \hline 
   MD, bare probe (Ref.~\cite{hormann2020sds}) $F_d\:(\text{nN})$              &  1.27 & 0.15 & 0.02 \\ 
   MD, covered probe $F_d\:(\text{nN})$              &  1.41 & 0.38 & 0.18  \\
   \hline 
\end{tabular}}
}
\caption {\label{tab:results_stokes_drag} Viscous drag in TIP3P water 
as estimated for a bead of $r = 2.5\: \text{nm}$ by Stokes' law and
as recorded via MD for our AFM tip model in bulk solution.
}
\end{table}

\begin{figure*}
  \includegraphics[width=372pt]{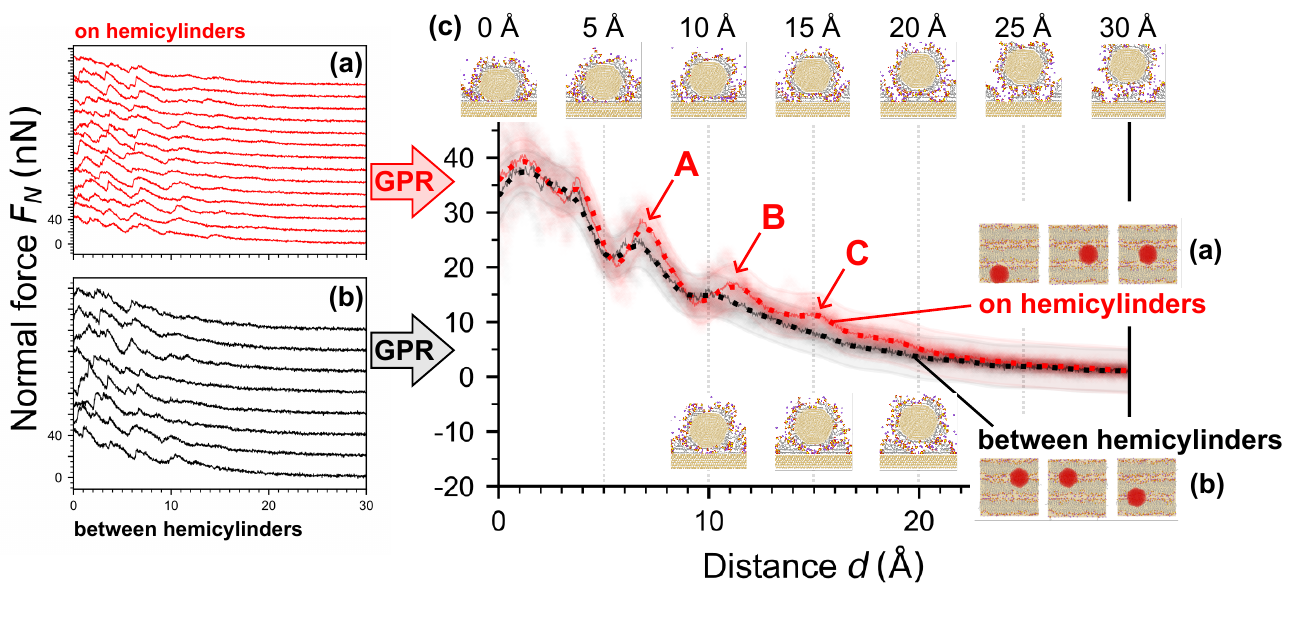}
  \caption{
  \textbf{(a)} 12 AFM force-distance curves recorded at varying sites with probe model centered above hemicylinder apex (red), \textbf{(b)} 8 AFM force-distance curves recorded at varying sites with probe model centered between two hemicylinders (black). Curves are displayed at a $20 \mathrm{nN}$ vertical spacing.
  \textbf{(c)} Gaussian process regression on all valid 2-ps averaged normal force - distance data points from set of trajectories. 
  Top-view insets labeled \textbf{(a)} and \textbf{(b)} as well show exemplary initial configurations  for probing sites on and between hemicylinders.
  The model for probing sites at the hemicylinders' apex is colored red, while the model for sites exactly between two hemicylinders is shown in black. A thin continuous line visualizes the simple mean of the 2-ps-averaged data points. Strong dots show the inferred Gaussian process mean. The translucent bands mark the $95 \%$ confidence interval for the models. Underlying data points are indicated as transparent coloring with varying intensity. Upper snapshots display exemplary cross-sections at normal intervals of $5\, \mathrm{\text{\AA}}$ of a probe approach on hemicylinders, lower snapshots show a probe coming down between two hemicylinders at $20$, $15$ and $10\, \mathrm{\text{\AA}}$. Red arrows point to features A - C present in the fit on trajectories with the probe model centered above hemicylinders, but absent from the fit for trajectories with the probe model centered between hemicylinders. Note that zero normal distance corresponds to the hypothetical exact overlap of outer probe and substrate gold layers in the absence of all interactions.}
  \label{fig:gpr_on_afm_force_distance_curves_on_and_between_hemicylinders}
\end{figure*}

All of the following results use an approach velocity of $1\,\text{m}\,\text{s}^{-1}$, which is a compromise between reduce solvent drag and computational cost.
Figures~\ref{fig:gpr_on_afm_force_distance_curves_on_and_between_hemicylinders}a and b show multiple approaches on hemicylinders.
Each force-distance curve was recorded at a different lateral offset of the tip.
More specifically, panel a shows results for approach of the probe on top of the hemicylinders, while panel b shows results between hemicylinders.
Top-down views of select configuration can be found in the insets to Fig.~\ref{fig:gpr_on_afm_force_distance_curves_on_and_between_hemicylinders}~c.
Single normal approach curves in panel a and b clearly exhibit distinct features.
As with the velocity-dependent force-distance curves of Fig.~\ref{fig:afm_force_distance_curves_velocity}, it is difficult to identify a systematic dependence of the individual features on the site (on or between hemicylinders) of approach.

We use stochastic variational Gaussian process regression~\cite{hensman2013gaussian}, a variant of Gaussian process regression (GPR)~\cite{krige1951statistical} suitable for large amounts of data, to fit trend-lines to all force-distance curves in panels a and b. 
Figure~\ref{fig:gpr_on_afm_force_distance_curves_on_and_between_hemicylinders} c shows the fit to the data shown in Fig.~\ref{fig:gpr_on_afm_force_distance_curves_on_and_between_hemicylinders} a and b.
Translucent bands indicate 95 \% confidence intervals.
Insets at the top of Fig.~\ref{fig:gpr_on_afm_force_distance_curves_on_and_between_hemicylinders} c show $5 \mathrm{\text{\AA}}$-spaced side views of one representative approach on top of hemicylinders along a cross-section as indicated in Fig.~\ref{fig:perspective-view} b.
Insets at bottom show three such snapshots for an approach between hemicylinders.

While single trajectories such as those shown in Fig.~\ref{fig:afm_force_distance_curves_velocity} do not necessarily provide generalizable insight, the averages over many trajectories (obtained from GPR) in Fig.~\ref{fig:gpr_on_afm_force_distance_curves_on_and_between_hemicylinders} filter out fluctuations and emphasize recurring features.
The approach on top of hemicylinders clearly shows several realignment or squeeze-out events absent from the approach between hemicylinders.
This is most obvious for the squeeze-out of multiple down to three molecular layers in the contact taking place around a normal distance of $15\,\mathrm{\text{\AA}}$ (arrow C) and the transition from three to two molecular layers in the contact around $11\,\mathrm{\text{\AA}}$ (arrow B).

The cross-sectional snapshots in Fig.~\ref{fig:gpr_on_afm_force_distance_curves_on_and_between_hemicylinders} clearly show the absence of these transitions in the case of the probe coming down between two hemicylindrical aggregates.
Below a normal distance of $10\,\mathrm{\text{\AA}}$, the force-distance curves do not show  distinguishable features anymore.
This corresponds to what is visible in the snapshots:
In both cases, the contact is filled with two molecular layers at $10\,\mathrm{\text{\AA}}$, undergoes the squeeze-out to one molecular layer around $6-7\,\mathrm{\text{\AA}}$ (arrow A), followed by the compression of probe and substrate around a few possibly trapped molecules or on direct gold-gold contact. 
Our GPR model shows the features A-C in Fig.~\ref{fig:afm_force_distance_curves_velocity} are characteristic for the morphology, while features E and F appear to be random features of the specific realization of the film.

Such squeeze-out events are representative for AFM experiments on surfactant films~\cite{burgess1999direct} or other confined liquids~\cite{krass2018molecular}.
For well-prepared molecularly smooth convex surfaces, this oscillatory behavior due to liquid layering and thin-film structure is observable even in macroscopic surface force apparatus (SFA) experiments~\cite{israelachvili1988forces}, but naturally this is not the case for rough surfaces~\cite{oshea2010liquid, israelachvili1991intermolecular}.

%

\begin{figure}
  \includegraphics[width=240pt]{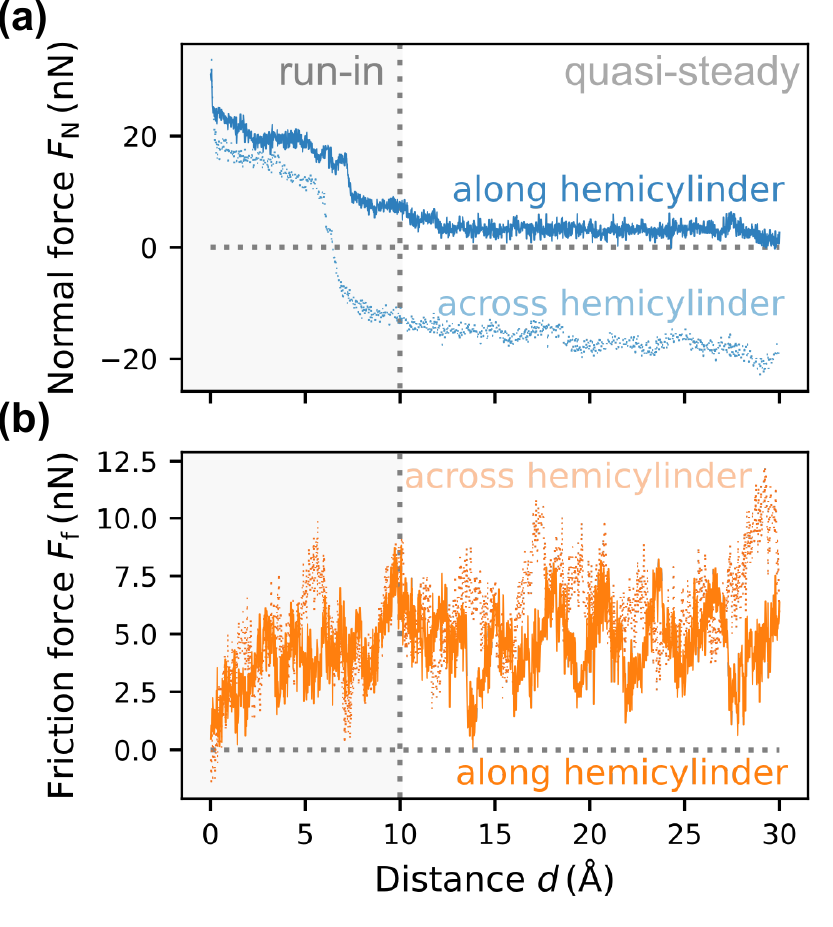}
  \\
  \caption{Typical \textbf{(a)} sample normal force (blue) and \textbf{(b)} friction force (orange) signals for sliding across (dotted line) and along (solid line) a hemiylindrical adsorption aggregate at fixed normal distance $3\,\mathrm{\text{\AA}}$ and $1\,\mathrm{m}\,\mathrm{s}^{-1}$ velocity. In the case of sliding across the hemicylindrical aggregate, a gold-gold contact occurs beyond $5\,\mathrm{\text{\AA}}$ lateral distance, resulting in adhesive normal force.}
  \label{fig:distance_normal_force_distance_friction_force_along_and_across_hemicylinders}
\end{figure}

We now focus on lateral sliding on these hemicylindrical aggregates.
Sliding is carried out at constant height, and we separately show normal force (Fig.~\ref{fig:distance_normal_force_distance_friction_force_along_and_across_hemicylinders}a) and friction force (Fig.~\ref{fig:distance_normal_force_distance_friction_force_along_and_across_hemicylinders}b) as a function of the sliding distance $d$ for a total distance of $3\,\text{nm}$.
To probe a potential directional dependence of the anisotropic hemicylinders, we run these calculations across (solid lines in Fig.~\ref{fig:distance_normal_force_distance_friction_force_along_and_across_hemicylinders}a and b) and along (dotted lines) the hemicylindrical aggregates.
Specifically, Fig.~\ref{fig:distance_normal_force_distance_friction_force_along_and_across_hemicylinders} shows results obtained at a fixed normal distance of $3\, \mathrm{\text{\AA}}$ of the tip.
The normal force, initially slightly higher for sliding along hemicylinders drops abruptly in both cases between a sliding distance of $5$ and $10\,\mathrm{\text{\AA}}$.
It stabilizes towards a steady-state around $5\,\mathrm{nN}$ for sliding along the hemicylinders, but drops into the adhesive regime for sliding across.
The friction force is, on average, slightly higher for sliding across than sliding along hemicylinders.
In contrast to the normal force, the friction force shows larger fluctuations and increases during the initial run-in period.

In our setup of fixed gap height and instantaneous onset of linear motion, we expect that the initial force response is dominated by the onset of sliding before approaching a (quasi) steady-state.
This behavior is observable in all our trajectories, and in extreme cases we observe the formation of a bare gold-gold contact, as shown by the dotted line with negative normal force in Fig.~\ref{fig:distance_normal_force_distance_friction_force_along_and_across_hemicylinders}a.
%
%
In the following analysis, we deliberately exclude all trajectories forming gold-gold contacts and label the first $1\,\mathrm{nm}$ of sliding as \emph{run-in}.
The subsequent $2\,\mathrm{nm}$ are labeled as \emph{quasi-steady}.
%
All our simulations show similar phenomenology in that we can identify separate run-in and quasi-steady sliding regimes.
%
%
However, our lateral sliding distance in single runs barely exceeds the radius of one hemicylindrical aggregate.
Consequently, our friction force results are never steady state across a periodically repeating pattern and our run-in regime should not be confused with run-in observed for macroscopic contacts~\cite{blau2005nature}. 

\begin{figure*}
  \includegraphics[width=0.7\textwidth, valign=t]{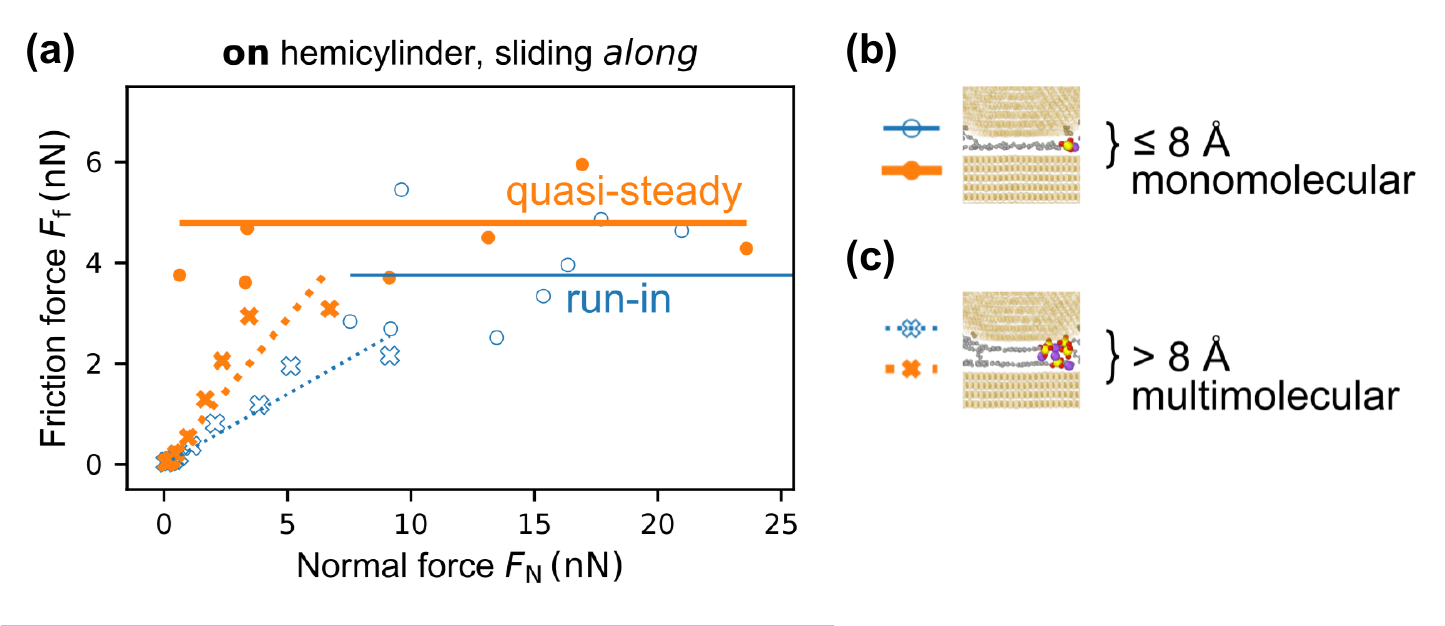}
  \\
  \caption{\textbf{(a)} As illustrated by blue and orange coloring in the inset repeating the force signal of Fig.~\ref{fig:distance_normal_force_distance_friction_force_along_and_across_hemicylinders}, we label the initial $1 \mathrm{nm}$ stretch \emph{run-in}, the subsequent stretch \emph{quasi-steady}, compute friction and normal force averages for both such stretches and show these friction force - normal force pairs as blue and orange data points. 
  Furthermore, we classify trajectories by their normal probe substrate distance. \textbf{(b)} At normal distances $z \leq 0.8\,\mathrm{nm}$, we speak of a gap of monomolecular height and show data points as circles. The mean friction force of these data points is indicated by a solid horizontal line in \textbf{(a)}. \textbf{(c)} At normal distances $z > 0.8\,\mathrm{nm}$, we speak of a gap of multimolecular height and show data points as crosses in \textbf{(a)}. Zero intercept linear fits to these data points are shown as dotted lines.}
  \label{fig:friction_force_normal_force_hemicylinders}
\end{figure*}

In the following, we analyze our friction results in these two categories.
Specifically, we average normal and friction force separately over the run-in and quasi-steady sliding regimes.
Figure~\ref{fig:friction_force_normal_force_hemicylinders}a is a scatter plot of the resulting average friction force versus normal force for all our runs on hemicylinders.
%
%
%
For both run-in and quasi-steady sliding, the data evidently exhibits two distinct friction regimes: linear Amontons friction at lower normal forces, and friction that is nearly independent of normal force at higher loads.
The transition from Amontons to constant force correlates with the transition from two or more molecular layers confined in the contact (see Fig.~\ref{fig:friction_force_normal_force_hemicylinders}b for a snapshot) down to only one molecular layer (Fig.~\ref{fig:friction_force_normal_force_hemicylinders}c).
This transition takes place below a normal distance of about $8\,\mathrm{\text{\AA}}$.
Consequently, we categorize data points into \emph{multimolecular} and \emph{monomolecular}, as indicated by crosses and solid symbols, respectively, in Fig.~\ref{fig:friction_force_normal_force_hemicylinders}a.
The figure also shows corresponding linear fits to the (multimolecular) Amontons regime as a dotted line, that explictly goes through the origin.
We additionally fit a constant to the monomolecular regime, shown as a solid line.

%

\begin{figure}
  \includegraphics[width=240pt]{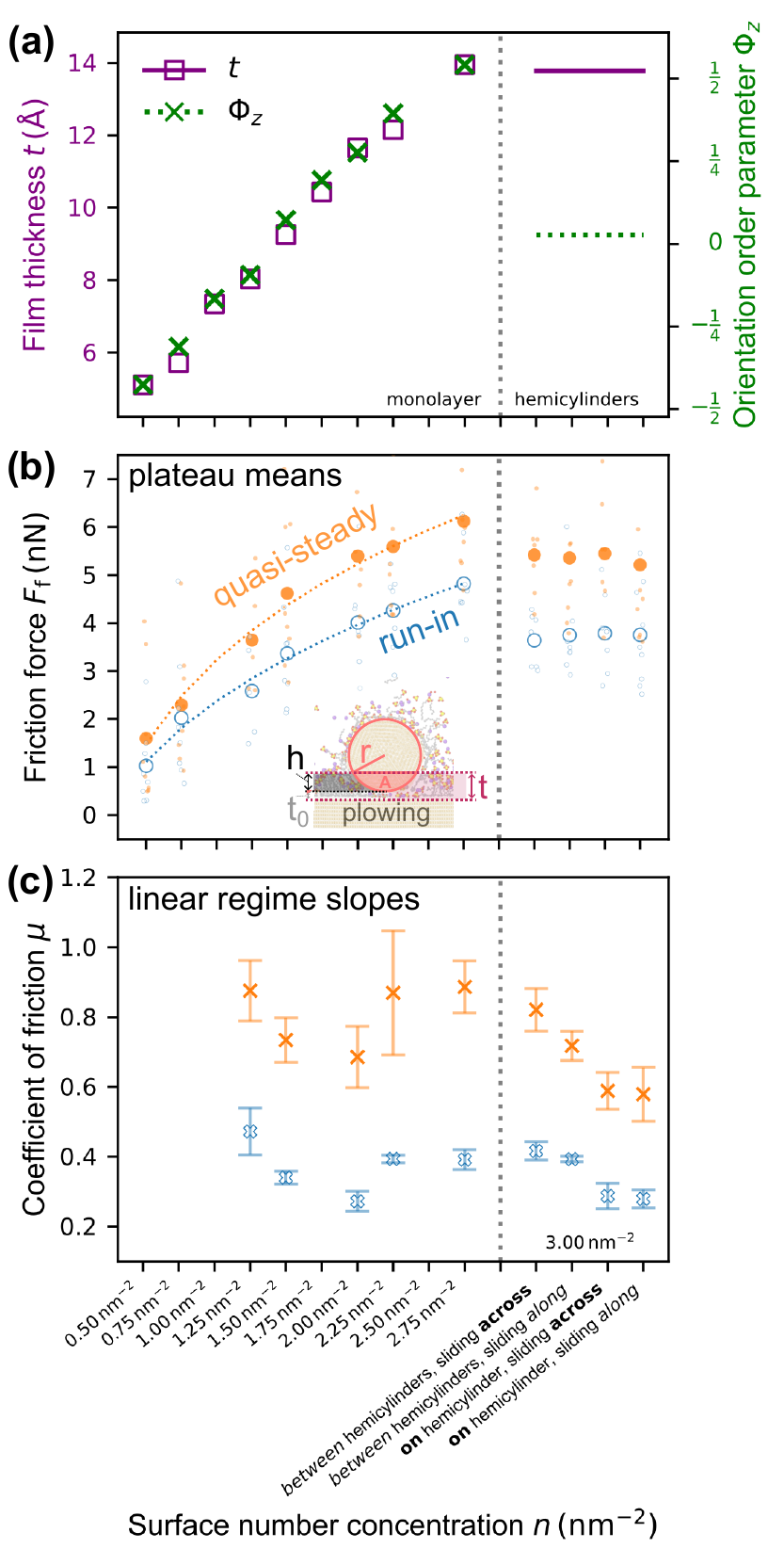}
  \\
  \caption{
  \textbf{(a)} Thickness $t$ (purple squares and solid line) and orientation order parameter $\Phi_z$ (green crosses and dashed line) of undisturbed films at increasing surface concentration. 
  \textbf{(b)} Narrow gap friction force means (together with the underlying plateau data points visible as lightly colored circles) for monolayer concentrations up to to $2.75\,\mathrm{nm}^{-2}$ and for hemicylindrical aggregates at $3\,\mathrm{nm}^{-2}$ distinguished by site and sliding direction. 
  Dashed fits in the monolayer regime of functional form $a + b\,x^c$ serve to guide the eye.
  The inset illustrates film thickness $t$ and plowing depth $h$. 
  \textbf{(c)} Wide gap coefficients of friction for monolayer concentrations up to to $2.75\,\mathrm{nm}^{-2}$ and for hemicylindrical aggregates at $3\,\mathrm{nm}^{-2}$ distinguished by site and sliding direction. 
  Markers adhere to the scheme introduced in Fig.~\ref{fig:friction_force_normal_force_hemicylinders}b and c;
  as before, blue and orange data points correspond to data from the run-in and quasi-steady regime.}
  \label{fig:friction_force_and_cof_vs_concentration_or_site_and_sliding_direction}
\end{figure}
\begin{figure*}
  \includegraphics[width=460pt, valign=t]{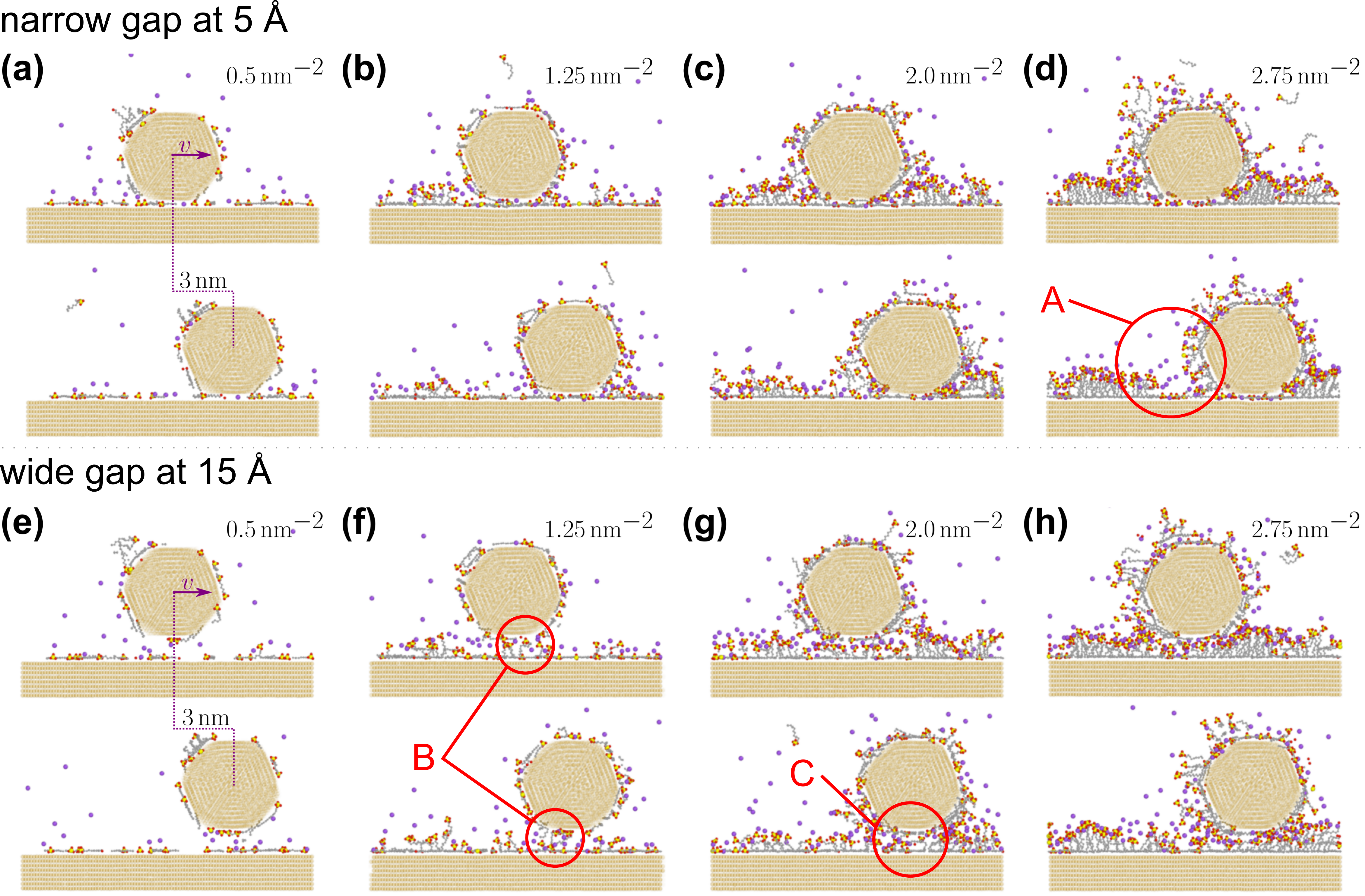}
  \\
  \caption{
  Cross-section snapshots of trajectories at fixed normal surface-surface distance \textbf{(a-d)} $5\,\mathrm{\text{\AA}}$ and \textbf{(e-h)} $15\,\mathrm{\text{\AA}}$ before (top) and after (bottom) lateral sliding for $3\,\mathrm{nm}$ at velocity $v = 1\,\mathrm{m}\,\mathrm{s}^{-1}$ on monolayers of surface concentrations \textbf{(a,e)} $0.5\,\mathrm{nm}^{-2}$, \textbf{(b,f)} $1.25\,\mathrm{nm}^{-2}$, \textbf{(c,g)} $2.0\,\mathrm{nm}^{-2}$, and \textbf{(d,h)} $2.75\,\mathrm{nm}^{-2}$. Feature A shows a groove opening up in the track of the probe when plowing into a dense film. Feature B shows monomers adhering to both probe and substrate for sparse adsorption films. Feature C shows a well-formulated boundary lubrication film without any such bridging chains.
  }
  \label{fig:illusrtative_snapshot_series_monolayer_far_distance}
\end{figure*}
\begin{figure*}
  \includegraphics[width=504pt, valign=t]{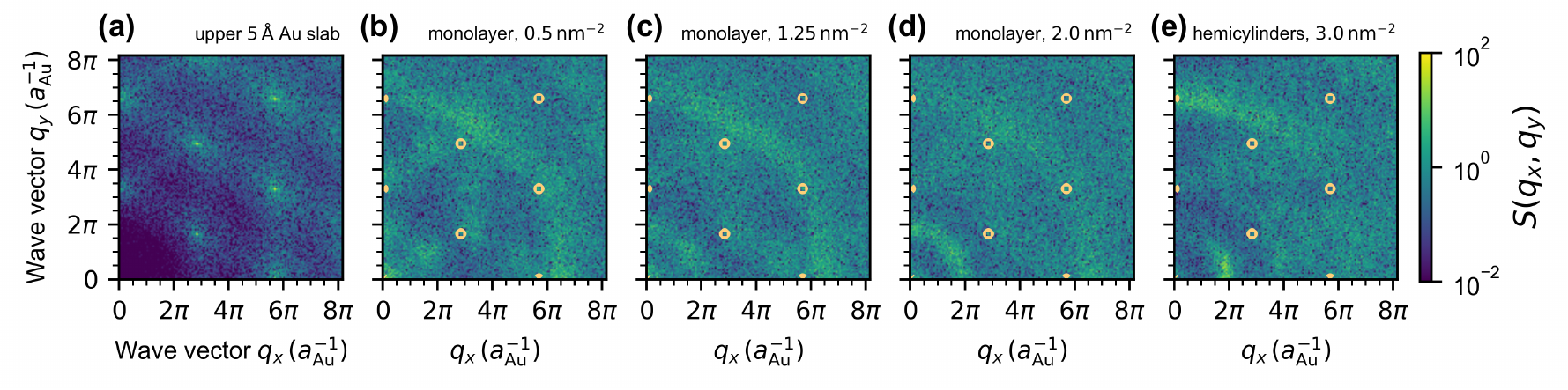}
  \\
  \caption{
  Two-dimensional structure factor $S(q_x, q_y)$ with wave vector $q_x$ and $q_y$ in units of the reciprocal gold lattice constant $a_\mathrm{Au}^{-1}$ on \textbf{(a)} the gold substrate's atomic layers in an upper $5\,\mathrm{\text{\AA}}$ slab and on the $\mathrm{DS}^{-}$ ion's carbon atoms in the single atomic layer adjacent to the gold substrate for \textbf{(b-d)} select monolayers of increasing concentration, and for \textbf{(e)} hemicylinders. Yellow circles in panels \textbf{(b-e)} repeat the structure factor peaks of the underlying gold substrate in panel \textbf{(a)}.
  }
  \label{fig:structure_factor_2d}
\end{figure*}

We have so far focused our detailed analysis on hemicylinders. Now, we extend our considerations onto monolayers.
%
%
In Fig.~\ref{fig:friction_force_and_cof_vs_concentration_or_site_and_sliding_direction}a we show the film thickness $t$ and orientation order parameter $\Phi_z$. 
We estimate the film thickness $t$ as the mean normal distance between the gold substrate's first atomic layer and the head groups of dodecyl sulfate ions in the adsorption film. 
The orientation order parameter $\Phi_z = \left(3 \left<\cos^2 \chi_z\right> - 1\right)/2$~\cite{seelig1974dynamic} is defined by the the angle $\chi_z$ between the substrate's surface normal $\hat{z}$ and the gyration tensor-based long principal axis of the dodecyl sulfate ion. In our case, the angular brackets indicate the mean over all molecules in the adsorption film. 
A value of nearly $-\frac{1}{2}$ at $0.5\,\mathrm{nm}^{-2}$ surface concentration indicates anti-alignment with $\hat{z}$. 
All monomers lie flat on the substrate.
Values above $0$ indicate partial alignment with $\hat{z}$.
At high concentrations, the monomers tilt upwards.
A value of $0$ indicates a uniform distribution of alignment.
That is the case for the fanned chains in hemicylindrical aggregates.
While $t$ and $\Phi_z$ show corresponding linear behavior with increasing concentration on monolayers, snapshots in Fig.~\ref{fig:illusrtative_snapshot_series_monolayer_far_distance} and the two-dimensional structure factor $S$ evaluated on a thin layer of gold in Fig~\ref{fig:structure_factor_2d}a and on the substrate-adjacent layer of carbon atoms in the dodecyl sulfate ions in Fig~\ref{fig:structure_factor_2d} reveal an important change in film morphology with increasing concentration.
At $0.5\,\mathrm{nm}^{-2}$, Fig.~\ref{fig:structure_factor_2d}b reveals alignment of flat-lying $\mathrm{DS}^{-}$ with the Au(111) crystal lattice of panel a. 
Fig.~\ref{fig:illusrtative_snapshot_series_monolayer_far_distance}f-h directly show how chains align with $\hat{z}$ with increasing concentration.
Correspondingly, the characteristic hexagonal pattern of the gold substrate gradually disappears from the structure factor in Fig.~\ref{fig:structure_factor_2d}c and e.
For hemicyindrical aggregates, the anisotropic ordering of chains is clearly discernible in Fig.~\ref{fig:structure_factor_2d}f.

The frictional phenomenology described above for hemicylinders also holds for monolayers at lower surface concentration:
The dependence of friction force on normal force splits into linear and quasi-constant regime.
Again, we extract a constant friction force $F_\text{f}$ from the normal load-independent regime and Amontons friction coefficient $\mu$ from the linear friction regime for our monolayers in the surface concentration range from $0.5$ to $2.75\,\mathrm{nm}^{-2}$.
Fig.~\ref{fig:friction_force_and_cof_vs_concentration_or_site_and_sliding_direction}b shows the saturated mean $F_\text{f}$ at high loads and narrow gaps as circles.
The small dots show the individual values of the friction force that were averaged to obtain the mean $F_\text{f}$.
The friction force in the quasi-steady regime rises monotonically from $\sim 1\,\text{nN}$ to $\sim 5\,\text{nN}$ with increasing monolayer concentration.
Dashed lines in panel b show fits of the friction data to an empirical power-law $a + b\,x^c$ with fit parameters $a$, $b$ and $c$.
The run-in regime follows this trend, albeit at lower $F_\text{f}$.
Note that the plot also shows the results obtained for hemicylinders at a surface concentration of $3\,\mathrm{nm}^{-2}$, where we separately report friction force for the two sliding directions.

Hemicylinders show a slightly lower $F_\text{f}$ of $\sim 4.5\,\text{nN}$ than monolayers at the highest surface concentration.
When sliding on hemicylinders, the different starting configurations and sliding directions do not appear to systematically differ (see Fig.~\ref{fig:friction_force_and_cof_vs_concentration_or_site_and_sliding_direction}b).
This means, the orientation of the hemicylinders does not introduce anistropy into the frictional response.
If any, there is a slight systematic difference of mean quasi-steady friction force from sliding across to sliding along hemicylinders. 
Prominent is the drop of the plateau friction force from dense monolayer at $2.75\,\mathrm{nm}^{-2}$ to hemicylindrical aggregates at $3\,\mathrm{nm}^{-2}$.

Similarly, Fig.~\ref{fig:friction_force_and_cof_vs_concentration_or_site_and_sliding_direction}c shows the friction coefficient $\mu$ at low normal forces, where Amontons' friction law holds.
The friction coefficient $\mu$ does not have a clear monotonic trend.
In the quasi-steady regime (filled markers in Fig.~\ref{fig:friction_force_and_cof_vs_concentration_or_site_and_sliding_direction}c), we see a decrease of $\mu$ from $0.9$ at $1.25\,\mathrm{nm}^{-2}$ to a minimum of $0.7$ at $2\,\mathrm{nm}^{-2}$, followed by a subsequent increase back to $0.9$.
For the hemicylindrical configurations, we observe $\mu$ of $0.8$ and $0.7$ for sliding across and along hemicylinders for an initial placement of the probe in a valley between two aggregates.
For an initial placement centered on the apex of an aggregate, $\mu$ decreases to $0.6$.
A qualitatively similar behavior is observable for the run-in regime (unfilled markers in Fig.~\ref{fig:friction_force_and_cof_vs_concentration_or_site_and_sliding_direction}c), with overall lower friction coefficients.
During run-in, the values of $\mu$ vary between $0.3$ to $0.5$ in the monolayer case.
For hemicylinders, it is $\sim 0.4$ for a probe initially placed between hemicylinders and $\sim 0.3$ for initial placement on a hemicylindrical aggregate.
%

Figure~\ref{fig:illusrtative_snapshot_series_monolayer_far_distance} shows snapshots of the initial and final configurations for a few select sliding trajectories which underlie the presented analysis.
Panels a-d show sliding at a narrow gap of $5\,\mathrm{\text{\AA}}$, where only a single molecular layer is confined between probe and substrate.
The surface coverage increases from left to right.
Such trajectories are representative for cases where Amontons' law does not hold.
Figures \ref{fig:illusrtative_snapshot_series_monolayer_far_distance}e-f show sliding at a wider gap of $15\,\mathrm{\text{\AA}}$, where multiple molecular layers can be accommodated between probe and substrate. 
Such trajectories are representative for Amontons' friction regime.

\section{Discussion}


The term interfacial friction has been suggested to distinguish the sliding of two molecularly smooth, undamaged surfaces, possibly lubricated by one or two molecular layers, from boundary lubrication, which implies plastic deformation and asperity contact on rough surfaces~\cite{israelachvili1992adhesion}.
Clearly, the sliding of atomically smooth surfaces of tip and substrate, lubricated by individual flat lying molecules of varying surface concentration at narrow gaps, falls into the category of interfacial friction.
Other molecular dynamics simulations consistently show Amontons behavior in the interfacial friction regime~\cite{he1999adsorbed}.
In all our simulations, friction forces become independent of normal forces at narrow gaps that accommodate a single molecular layer at most.
At such close distances, there is a competition of adhesive gold-gold interaction~\cite{grochola2005fitting} and repulsion by compressed substrate and confined molecules.
This leads to a wide range of net normal forces across the narrow range of surface-surface distances below $8\,\mathrm{\text{\AA}}$ with little variation in mean lateral forces.

Friction felt by the probe must be made up by viscous drag in the solvent, plowing into the surfactant film, and \emph{cobblestone}-like sliding on the atomistic scale, where the atomic granularity of gold surface lattices and trapped surfactant molecules plays significantly into friction~\cite{homola1989measurements}. 
Friction forces oscillate at the order of $\mathrm{\text{\AA}ngstrom}$ during lateral sliding in Fig.~\ref{fig:distance_normal_force_distance_friction_force_along_and_across_hemicylinders}b, indicative of the cobblestone mechanism and pointing to the Au(111) facets' lattice with a  nearest neighbor distance of $\sim 2.88\,\mathrm{\text{\AA}}$. 
The friction force in Fig.~\ref{fig:concept-afm-approach}b shows that these oscillations disappear at surface-surface distances larger than a few $\mathrm{\text{\AA}ngstrom}$.

In Fig.~\ref{fig:friction_force_and_cof_vs_concentration_or_site_and_sliding_direction}b, we observe a sublinear friction force vs. concentration relationship towards the upper end of the investigated concentration range at narrow gaps. 
%
%
At sparse surface coverage with flat-lying monomers (concentrations below $1\,\mathrm{nm}^{-2}$), the probe moves across the film laterally only subject to the viscous drag of water and the discussed cobblestone contribution.
For higher concentrations, the probe has indented the surfactant aggregates normally and plows into the adsorption film laterally, opening up a groove at its tail as illustrated by feature marker A in Fig.~\ref{fig:illusrtative_snapshot_series_monolayer_far_distance}d.
\hemicylinders{
The structure factor of substrate-adjacent monomers in Fig.~\ref{fig:structure_factor_2d}e show that in the case of hemicylindrical aggregates, many monomers are lying flat on the gold substrate, not unlike the flat-lying monolayers at lower concentrations, but well aligned along the cylinder axis.
They form the foundation of hemicylindrical domes assembled above.
At narrow gaps, the probe slides across this foundational monolayer while simply peeling off the surfactant domes.
}
Snapshots of Fig.~\ref{fig:illusrtative_snapshot_series_monolayer_far_distance}g and h and the absence of any ordering in the structure factors of Fig.~\ref{fig:structure_factor_2d}d and e reveal that there is no such foundational layer of flat-lying monomers in case of the densely packed monolayer, and consequently no simple peel-off is possible.

Since we have not investigated the velocity dependence in our lateral sliding simulations, we cannot decide to what extent the plowing probe experiences velocity-dependent viscous drag or velocity-independent resistance due to elastoplastic deformation of the film.  
Experiments have shown SDS adsorption films to start behaving viscous around a surface coverage of $2\,\mathrm{nm}^{-2}$~\cite{poskanzer1975surface}.
In the earliest plowing models for soft metals, friction force $F_\text{f} = F_\text{s} + F_\text{p}$
is the sum of $F_\text{s}$, the force required to shear the metallic junctions, and $F_\text{p} = A p_\text{Y}$, the force required to displace the softer metal from the path of the slider~\cite{bowden1943ploughing}, where $A$ is the cross section area of the plowing track and $p_\text{Y}$ the pressure to cause plastic flow of the softer metal, its indentation hardness.
We use this simple model for an order-of-magnitude estimate of the adsorption film's flow pressure $p_\text{Y}$. 
The plowing depth amounts to $h = t - t_0$ as illustrated by the inset in Fig.~\ref{fig:friction_force_and_cof_vs_concentration_or_site_and_sliding_direction}a. 
Neglecting viscous drag in water, we relate the discussed cobblestone contribution in our systems to $F_\text{s}$ and the displacement of surfactant molecules in the plowing track to $F_\text{p}$. 
This approach leads to an estimate mean $p_Y$ at the order of $1\,\mathrm{GPa}$ for the range from $1.0$ to $2.75\,\mathrm{nm}^{-2}$.

%
For wider gaps, plowing friction plays a reduced to vanishing role. 
Gold-gold adhesion and cobblestone oscillations disappear. 
We enter a regime of well-behaved boundary lubrication where the adsorption films on probe and substrate remain largely intact, as seen in Fig.~\ref{fig:illusrtative_snapshot_series_monolayer_far_distance}e-h. 
Studies on bare probes penetrating chemisorbed monolayers~\cite{summers2017investigating, gao2021scaledependent} or symmetric boundary-lubricated flat-on-flat systems~\cite{eder2013derjaguin} report adhesive effects and introduce the Derjaguin offset $F_0$ as a modification to Amontons' law, $F_\text{f} = \mu\,F_\text{N} + F_0$.
In our model of probe and substrate covered with physisorbed anionic surfactant molecules, we have not observed any net adhesive interaction in the normal approach force-distance curves. 
Accordingly, none of our friction force--normal force relations exhibits a considerable finite intercept.
We observe Amontons friction $F_\text{f} = \mu\,F_\text{N}$, as illustrated with cross markers and dashed linear fits in Fig.~\ref{fig:friction_force_normal_force_hemicylinders}.

Other than the monotonically increasing saturated friction forces at narrow gaps in Fig.~\ref{fig:friction_force_and_cof_vs_concentration_or_site_and_sliding_direction}b, the concentration-dependent friction coefficient at wide gaps in Fig.~\ref{fig:friction_force_and_cof_vs_concentration_or_site_and_sliding_direction}c clearly shows a minimum for both run-in and quasi-steady regime in the concentration range of $1.5$ to $2.0\,\mathrm{nm}^{-2}$ for a monolayer.
To understand the minimum in the monolayer case, we discuss Fig.~\ref{fig:illusrtative_snapshot_series_monolayer_far_distance}f-h together with the corresponding friction coefficients at concentrations $1.25$, $2.0$, and $2.75\,\mathrm{nm}^{-2}$ in Fig.~\ref{fig:friction_force_and_cof_vs_concentration_or_site_and_sliding_direction}c.
In Fig.~\ref{fig:illusrtative_snapshot_series_monolayer_far_distance}f at $1.25\,\mathrm{nm}^{-2}$, sparse coverage with flat-lying monomers at probe and substrate may introduce 
single monomers bridging the gap, emphasized by feature marker B.
In Fig.~\ref{fig:illusrtative_snapshot_series_monolayer_far_distance}g at $2.0\,\mathrm{nm}^{-2}$, denser coverage leads to a well-formulated lubrication layer pointed out by marker C.
%
In Fig.~\ref{fig:illusrtative_snapshot_series_monolayer_far_distance}g at $2.75\,\mathrm{nm}^{-2}$, very dense coverage introduces a slight plowing contribution again. 

For hemicylinders, the friction coefficient shows a clear site dependence. 
Beginning to slide on a hemicylindrical aggregate yields lower friction than beginning to slide in the middle between two aggregates.
This low friction coefficient reaches the monolayer minimum, and even exceeds it in the case of the quasi-steady regime.
To explain the difference of friction coefficients for an initial configuration centered between two hemicylindrical aggregates and a starting configuration on top of an aggregate, we rely on similar arguments.
Former starting position means a groove for the probe to embed in and an increased likelihood of hydrophobic attraction between hydrocarbon tails, similar to the case of low monolayer concentrations.
Latter starting position means densely packed monomer layering to bear the probe, similar to the case of higher monolayer concentrations.

In macroscopic friction tests, higher adsorption mass is usually associated with improved lubrication.
This appears to be the case for SDS adsorption films on different base materials~\cite{liu2022online,liu2021active,zhang2015boundary, zhang2014stick}, that show decreased kinetic friction coefficient with increasing adsorption mass as long as the film does not yet transition to the hemicylinder phase.
In our microscopic model with a sharp probe, however, we observe exactly the inverse of this trend for the normal force-independent response at narrow gaps in Fig.~\ref{fig:friction_force_and_cof_vs_concentration_or_site_and_sliding_direction}b.
Consistent with our observation, Nalam et al.~\cite{nalam2019adsorption} have experimentally shown the increase of friction with increasing adsorption film density for fatty acids probed with a sharp AFM tip.
Similarly, Gao et al.~\cite{gao2021scaledependent} have elucidated the relation between probe curvature, film density, and friction under plowing into such organic friction modifier (OFM) monolayers using MD simulations.
The higher the concentration, and the sharper the probe, the more prominent the plowing effect and hence the higher the friction coefficient in constant normal force simulations.

In the regime of Amontons' friction, we observe a minimum in the concentration-dependent friction coefficient that arises due to the complex interplay between adhesive and repulsive normal forces, and plowing into the adsorption film.
A minimum in the concentration-dependent coefficient of friction on SDS adsorption films on stainless steel has indeed been observed in macroscopic ball-on-flat experiments~\cite{zhang2014stick, liu2023mitigation}.
In the image of Bowden's model for boundary lubrication~\cite{bowden1940friction, bowden1945lubrication}, a break down of lubricant film leads to metallic junctions that contribute to the friction force.
It has been argued that an SDS adsorption film in the monolayer phase prevents the formation of such junctions, improving lubrication with increasing concentration, while the transition to hemicylindrical stripe-like aggregates at again higher concentrations exposes bare metal substrate between the stripes and hence facilitates junction formation, promoting stick-slip~\cite{zhang2014stick}.

Typical ball-on-flat experiments, such as those of Ref.~\cite{zhang2014stick}, have rough interfaces and run at velocities on the order of $\mathrm{mm}\,\mathrm{s}^{-1}$.
Our simulations describe different spatial and time scales.
Still, the concentration range of the macroscopic friction minimum before the onset of stick-slip and of the nanoscopic minimum observed here coincide.
We believe latter contributes an asperity-level aspect to the macroscopic friction minimum so far unconsidered.

\section{Summary \& conclusion}

We showed that MD-simulated AFM force--distance curves and sliding measurements exhibit morphology-dependent characteristics.
For this purpose, we moved a $5\,\mathrm{nm}$ model gold AFM tip (or model asperity) normally onto or laterally along a $15\times 15\times 15\,\mathrm{nm}^3$ gold substrate block.
All probe facets and the contacting substrate surface were covered by SDS adsorption films of equivalent density to mimic a nanotribological system in the absence of any externally applied potential bias between the two contacting bodies that are facing each other with Au(111) planes.
Specifically, we investigated preassembled monolayers at surface concentrations concentrations from $0.5\,\mathrm{nm}^{-2}$ to $2.75\,\mathrm{nm}^{-2}$ and hemicylindrical aggregates at $3.0\,\mathrm{nm}^{-2}$.

In short, we have highlighted the impact of different parametric dimensions, namely probe velocity, surfactant concentration, film morphology and probing site on the characteristic features of force--distance curves.
In our lateral sliding simulations, we have been able to draw statistically reliable conclusions upon the manifestation of plowing friction and boundary lubrication on intact film layers in a nanotribological system by evaluation of a large number of trajectories.
Friction forces saturate around a nearly normal force-independent plateau at narrow gaps and high normal force.
At very close distances, a cobblestone-contribution induced by the Au(111) facets' atomic lattices arises in the friction signal.
In this regime, the probe plows deeply into the dense adsorption film.
In agreement with classical plowing models, this leads to increased friction with increased film density, an inversion of the expectation of improved lubrication with increased adsorption mass in macroscopic friction experiments expected for sharp asperities.
At intermediate probe-substrate distances and moderate normal load, our systems adhere to Amontons' law without any adhesive contribution.
Coefficients of friction range from $0.3$ to $0.9$, depending on morphology, site, and concentration.
The contact resembles a boundary-lubricated friction couple with intact adsorption layers within this regime.
The concentration-dependent monolayer friction coefficient reaches a minimum around $1.5$ to $2\,\mathrm{nm}^{-2}$, which we attribute to the complex interplay of hydrophobic attraction, the film's repulsion under compression, and the onset of plowing.
Similar mechanisms can explain the site and sliding direction dependency of the hemicylinders friction coefficient as well.

\section*{Acknowledgement}

We used \textsc{LAMMPS}~\cite{thompson2022lammps} for all molecular dynamics simulations.
Adsorption films were packed with \textsc{PACKMOL}~\cite{martinez2009packmol}.
Preparatory molecular dynamics runs have been carried out with \textsc{GROMACS}~\cite{abraham2015gromacs}.
Probe and substrate systems were merged with the help of \textsc{VMD}~\cite{humphrey1996vmd} and its \textsc{pbctools}, 
\textsc{topotools}~\cite{kohlmeyer2017akohlmey} and \textsc{mergetools} plugins.
Simulations runs were orchestrated with \textsc{Fireworks}~\cite{jain2015fireworks} and \textsc{dtool}~\cite{olsson2019lightweight}.
Rendered images have been produced with \textsc{OVITO}~\cite{stukowski2009visualization}.
Simulations were carried out on NEMO at the University of Freiburg (Deutsche Forschungsgemeinschaft grant INST 39/963-1 FUGG) and on JUWELS at the Jülich Supercomputing Center (projects hfr13 and hfr21).
Data is stored on bwSFS (University of Freiburg, Deutsche Forschungsgemeinschaft grant INST 39/1099-1 FUGG).


%

\end{document}